\documentclass[journal=jacsat,manuscript=article]{achemso}
\usepackage{caption}
\usepackage{float}
\usepackage{subcaption}
\usepackage{color}
\DeclareCaptionFormat{subfig}{\figurename~#1#2#3}
\DeclareCaptionSubType*{figure}
\usepackage[version=3]{mhchem} 

\author{Gengyue Dong}
\affiliation{Department of Materials, Imperial College London, South Kensington Campus, London SW7 2AZ, United Kingdom}

\author{Simão João}
\affiliation{Department of Materials, Imperial College London, South Kensington Campus, London SW7 2AZ, United Kingdom}

\author{Hanwen Jin}
\affiliation{Department of Materials, Imperial College London, South Kensington Campus, London SW7 2AZ, United Kingdom}

\author{Johannes Lischner}
\email{j.lischner@imperial.ac.uk}
\affiliation{Department of Materials, Imperial College London, South Kensington Campus, London SW7 2AZ, United Kingdom}
\alsoaffiliation{The Thomas Young Centre for Theory and Simulation of Materials, London E1 4NS, United Kingdom}

\title[An \textsf{achemso} demo]
  {Atomistic Theory of Hot-Carrier Generation in Aluminum Nanoparticles}

\abbreviations{IR,NMR,UV}

\begin{document}

\begin{tocentry}
    \includegraphics[height=4.43cm]{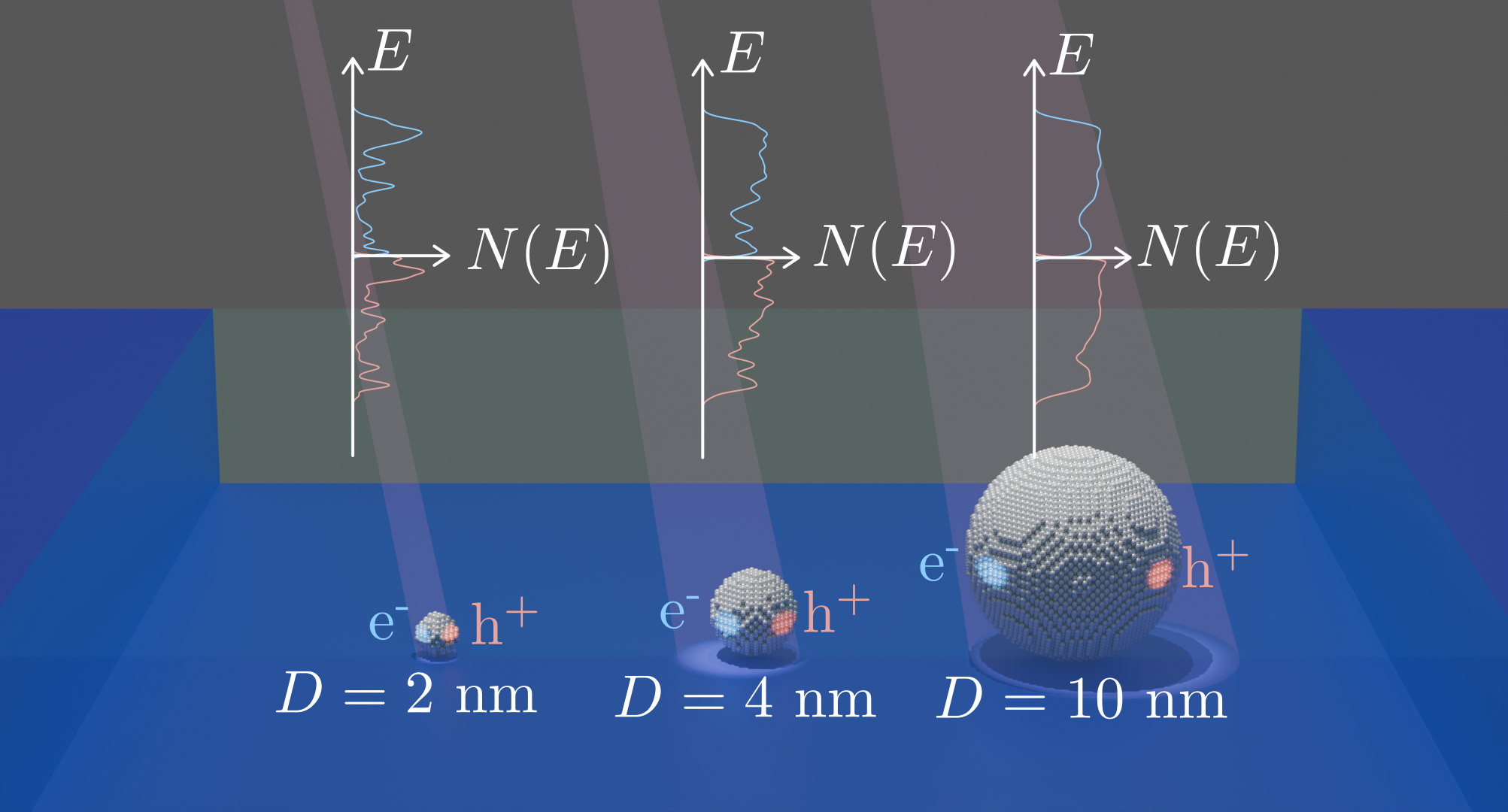}
\end{tocentry}


\begin{abstract}
Hot electrons and holes generated from the decay of localized surface plasmons (LSPs) in aluminum nanostructures have significant potential for applications in photocatalysis, photodetection and other optoelectronic devices. Here, we present a theoretical study of hot-carrier generation in aluminum nanospheres using a recently developed modelling approach that combines a solution of the macroscopic Maxwell equation with large-scale atomistic tight-binding simulations. Different from standard plasmonic metals, such as gold or silver, we find that the energetic distribution of hot electrons and holes in aluminium nanoparticles is almost constant for all allowed energies. Only at relatively high photon energies, a reduction of the generation rate of highly energetic holes and electrons close to the Fermi level is observed which is attributed to band structure effects suppressing interband decay channels. We also investigate the dependence of hot-carrier properties on the nanoparticle diameter and the environment dielectric constant. The insights from our study can inform experimental efforts towards highly efficient aluminum-based hot-carrier devices.

\end{abstract}
\newpage

\section*{Introduction}

The decay of localized surface plasmons (LSPs) in metallic nanoparticles generates energetic electrons and holes.\cite{govorov2006gold,hartland2017hotElectrons}. These hot carriers can be harnessed in nanoscale devcies for photodetection \cite{duan2019, kooKim2022Plasmonic, alamHotHoleTransfer, zhang2021Photoelectrochemical}, photocatalysis \cite{zhao2017photocatalysis, fujishima1972electrochemical, yan2016quantum, salvador2012electron,SimaospaperNatComms,HerranEmi,Wu2022,Chen2025}, and photovoltaics \cite{Polman2012Photonic,kim2018solar, renXuAdvancedHeterostructure, roudgarAmoliShariatiniaSynergisticInfluence, goniHotElectronGenerators,Jang2016Plasmonic,Chen2019Ultrathin}. In particular, current semiconductor-based devices for harvesting solar energy cannot absorb photons with energies smaller than the band gap.\cite{asfia2022pressureInfluence,islam2021semiconductorTransition,Pulfrey1974Schottky,Nelson2003Physics}. This limitation can be overcome by combining semiconductors with metallic nanoparticles which can absorb sub-bandgap photons and then inject hot carriers into the semiconductor thereby enhancing device efficiencies~\cite{garcia2013plasmoElectric,su2023hotCarrierPV}. Standard plasmonic metals, such as gold or silver, exhibit excellent optical properties \cite{ross2021low}, but their high cost restricts their use in large-scale devices. This challenge motivates the exploration of alternative plasmonic materials that are earth-abundant and therefore cheaper~\cite{gutierrez2018plasmonics,Sayed_2022}.

Aluminum, the third most abundant element in the earth's crust, holds substantial potential for large-scale applications due to its low cost \cite{Bonfiglio2023Aluminum,Salunkhe2023AluminumComposite}. Aluminum nanoparticles exhibit strong plasmon resonances that can be tuned from the ultraviolet into the visible spectrum \cite{atwater2010plasmonics}. This UV capability is a distinct advantage over gold and silver; for instance, Dubey et al. developed an aluminum plasmonics-enhanced GaN photodetector achieving record-high responsivity (670 A W$^{-1}$) and detectivity (1.48 × 10$^{15}$ cm Hz$^{\frac{1}{2}}$ W$^{-1}$) at 355 nm. Their work demonstrated aluminum's superior performance for UV applications due to its high plasma frequency and low intrinsic loss in this regime \cite{Dubey2020_AluminumPlasmonics}. Beyond optoelectronics, the Halas group's pioneering studies demonstrated the potential of aluminum nanostructures for hot-carrier generation and plasmonic photocatalysis, highlighting the importance of the native oxide on the nanoparticle surface for carrier extraction and catalytic performance \cite{Knight2014_ACSNano_AlForPlasmonics,Brongersma2015NatNano}. Subsequent studies from the same group introduced aluminum-based antenna–reactor architectures that exploit the tunability of the oxide shell to control hot-carrier transfer processes \cite{Bayles2024PNAS}. To bridge the gap to practical use, large-scale fabrication techniques using self-assembly nanoparticle template methods have also enabled the controllable preparation of aluminum nanoparticles with tunable sizes, demonstrating their potential for scalable hot-carrier devices \cite{Wang2023AluminumNanobowls}.

Besides experimental investigations, theoretical modeling has been critical for elucidating the fundamental mechanisms governing hot-carrier dynamics in aluminum nanostructures. Sundararaman et al.\cite{Sundararaman2014} and Zhang\cite{zhang2021} employed density functional theory (DFT) to study the generation of hot carriers from surface plasmon decay and their relaxation due to electron–electron and electron–phonon interactions in aluminum and other metals. Douglas et al. used a computational approach that combines reactive force field molecular dynamics with DFT\cite{C7NR04904H}. They simulated the oxidation of aluminum nanoclusters and analyzed the resulting structures using the time-dependent density functional tight-binding approach to reveal how oxidation affects plasmonic properties. Nordlander and coworkers employed electromagnetic simulations based on the finite-element method coupled with a Monte Carlo approach to investigate aluminum-based antenna-reactor photocatalytic systems\cite{robatjazi2017plasmon}. They demonstrated that the native aluminum oxide layer can function as a catalytically active reactor when paired with the plasmonic aluminum core. While these studies have provided important insights into the behaviour of hot carriers in aluminum nanoparticles, a detailed systematic understanding of the dependence of hot-carrier properties on nanoparticle size, photon energy and environment dielectric constant is still missing.

In this paper, we use a recently developed atomistic modelling technique which combines a solution of the macroscopic Maxwell equations with large-scale tight-binding models to evaluate Fermi's golden rule to study hot-carrier generation in aluminum nanoparticles (AlNPs).~\cite{jin2020plasmon} We present results for spherical AlNPs with diameters up to 10 nm which contain more than 30,000 atoms. In contrast to standard plasmonic materials, such Ag and Au, we find that the hot-carrier generation rates in AlNPs are almost constant as function of hot-carrier energy over the allowed energy range. Only for high photon energies, a reduction in the generation rate of low-energy electrons and high-energy holes is observed which is attributed to band structure effects affecting interband transitions. We also analyze the dependence of hot-carrier generation rates on the nanoparticle size and the dielectric constant of the nanoparticle environment. The insights from these calculations can inform experimental efforts towards highly efficient aluminum-based hot-carrier devices.

\section*{Methods}

\subsection{Absorption cross-section}
In the quasistatic approximation, the absorption cross-section of a spherical nanoparticle of radius $R$ embedded in an environment with dielectric constant $\epsilon_m$ is given by
\begin{equation}
    C_{\text{abs}}(\omega)  = \frac{8\pi^2}{\lambda} R^3 \text{Im} \left[ \frac{\epsilon(\omega) - \epsilon_m}{\epsilon(\omega) + 2\epsilon_m} \right],
    \label{eq:absorbption}
\end{equation}
where $\lambda$ is the wavelength of the light and $\epsilon(\omega)$ is the dielectric function of bulk aluminum taken from experiment.\cite{CRC2016} The absorption cross-section typically exhibits a peak at the LSP frequency, i.e., when $\epsilon(\omega_{\text{LSP}}) = -2\epsilon_m$. The quasistatic approximation is accurate when $\lambda \gg R$.

It can be seen from Eq.~\eqref{eq:absorbption} that the absorption cross-section can be tuned by embedding the nanoparticle in host media with different optical dielectric constants. Moreover, the oxide layer on the surface of the nanoparticle results in a redshift of the LSP frequency \cite{Chan2008_JPCC_Al_Triangles,Gerard2015_JPD_AlPlasmonics,Knight2014_ACSNano_AlForPlasmonics, Gutierrez2016_OE_OxideShellUV}. In our simulations, we treat $\epsilon_m$ as an adjustable parameter which captures all external factors, such as embedding medium, substrate and oxide shell, that affect the frequency of the localized surface plasmon.

\subsection*{Hot-Carrier Generation Rate}

The hot-electron generation rate \( N_e (E,\omega) \) per unit volume and energy of aluminum nanoparticles is given by
\begin{equation}
    N_e (E,\omega) = \frac{2}{V} \sum_{if} \Gamma_{if} (\omega) \delta(E - E_f),
\end{equation}
where \( V \) denotes the volume of the nanoparticle, and \( \Gamma_{if} \) is the transition rate between initial state $i$ and final state $f$ (with energies \( E_i \) and \( E_f \), respectively), induced by the total potential \( \hat{\Phi}_{\text{tot}} (\omega) \)  which includes the electric potential of the light and the induced potential resulting from the dielectric response of the nanoparticle. \( \Gamma_{if} \) is given by Fermi's golden rule~\cite{Manjavacas2014PlasmonInduced, DalForno2018PlasmonInduced}
\begin{equation}
    \Gamma_{if} (\omega) = \frac{2\pi}{\hbar} \left| \langle f | \hat{\Phi}_{\text{tot}} (\omega) | i \rangle \right|^2 \delta(E_f - E_i - \hbar\omega) f(E_i) (1 - f(E_f)),
\end{equation}
where \( f(E) \) denotes the Fermi-Dirac distribution function at room temperature and the total potential operator \( \hat{\Phi}_{\text{tot}}(\omega) \) is evaluated using the quasi-static approximation \cite{jin2020plasmon,jin2023}, which is valid because we only consider nanoparticles with diameters up to 10 nm, i.e. much smaller than the wavelength of light.

A tight-binding basis is used to represent the nanoparticle states \( |i\rangle \) and \( \langle f| \) in Eq.~(2). We assume that the wavefunctions of states that are involved in the LSP decay can be accurately represented as linear combinations of 3s, 3p, and 3d atomic orbitals. The corresponding tight-binding Hamiltonian is constructed using an orthogonal two-center parameterization derived from \textit{ab initio} density-functional theory calculations \cite{Papaconstantopoulos2015HandbookBandStructure}. To efficiently evaluate Fermi’s golden rule for large nanoparticles, we use the kernel polynomial method \cite{jin2020plasmon,Weisse2005KernelPolynomial,Joao2020KITE,JoaoLopes2018SpectralMethods}. To reduce the statistical error introduced by the stochastic trace evaluation of the kernel polynomial method we use a large number of random vectors: for small nanoparticle with diameters of 2 nm, 6000 random vectors are required to achieve convergence, while only 200 random vectors are needed for nanoparticles with a diameter of 4 nm and 10 nm. Similar techniques are used to calculate the electronic density of states (DOS) of the nanoparticles.

\section{Results and discussion}

We study the optical, electronic, and opto-electronic properties of spherical AlNPs. Nanoparticles with diameters of 2~nm (252 atoms), 4~nm (2,021 atoms), and 10~nm (31,575 atoms) are investigated. We also investigate the effect of different dielectric environments on the nanoparticle properties.

\begin{figure}[htbp]
    \centering
    \includegraphics[width=0.9\textwidth]{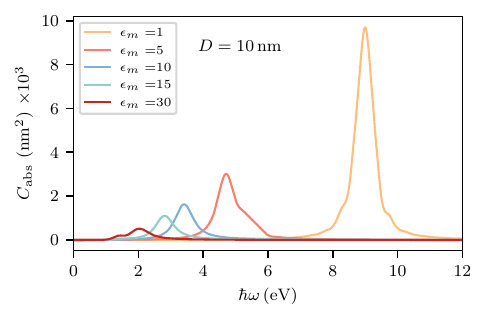} 
    \caption{Quasistatic absorption cross-sections $C_{abs}$ of spherical Al nanoparticles with 10 nm diameter, embedded in environments with dielectric constants $\epsilon_m=$ 1, 5, 10, 15, and 30.}
    \label{fig:hot_carrier_dielectric}
\end{figure}

\subsection*{Absorption cross-section in different environments}

Figure~\ref{fig:hot_carrier_dielectric} shows the absorption cross-sections of spherical AlNPs embedded in environments with different dielectric constants $\epsilon_m$. As $\epsilon_m$ increases, the absorption peak redshifts from the ultraviolet region to the visible spectrum. In vacuum, the peak is found at 9.0 eV, while for $\epsilon_m=30$ its energy is reduced to 2.0 eV. For $\epsilon_m = 30$, the absorption cross-section exhibits a pronounced shoulder near 1.4~eV, which originates from an optically active interband transition.\cite{vanDerVliet2016} In addition to the redshift of the LSP peak, a significant decrease in the LSP peak height is observed. This is caused by an increase in the imaginary part of $\epsilon(\omega)$ at low frequencies.\cite{knight-halas}

\subsection*{Electronic density of states}

\begin{figure}[htbp]
    \centering
    \includegraphics[width=0.7\textwidth]{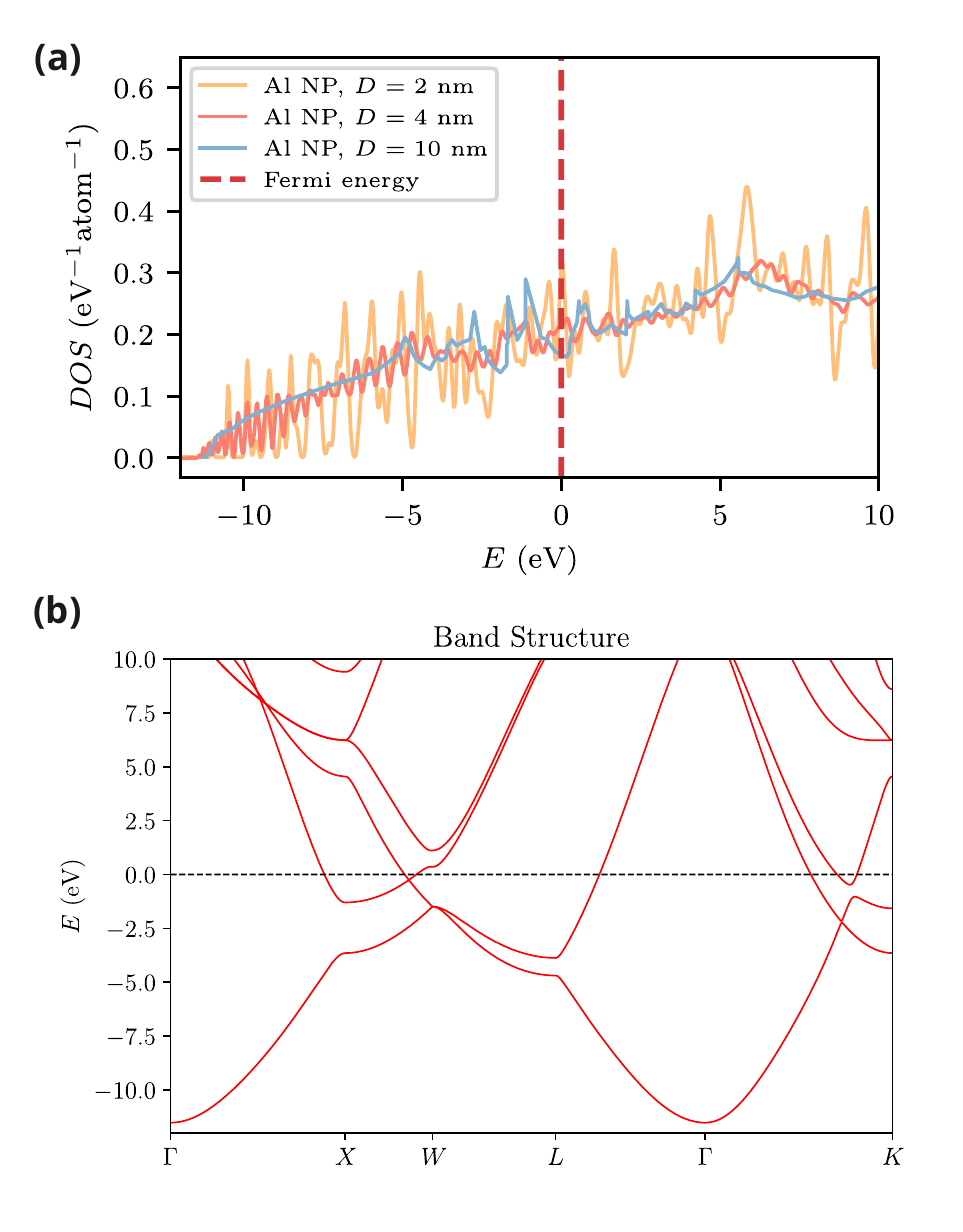} 
    \caption{\textbf{(a)} Density of states of spherical aluminum nanoparticles of diameters $D$ $=$ 2 nm, 4 nm, and 10 nm from tight-binding. \textbf{(b)} Band structure of bulk Al obtained from a tight-binding calculation.}
    \label{fig:dos}
\end{figure}

Figure~\ref{fig:dos}~(a) shows the density of states (DOS) of spherical AlNPs with different diameters. For the smallest nanoparticles ($D =2$~nm), the DOS is characterized by a series of sharp peaks that reflect the discreteness of the electronic states arising from quantum confinement effects. As the diameter increases, the discrete peaks gradually merge to form a continuous curve. At low energies, the nanoparticle DOS closely resembles that of a free electron gas, which is proportional to the square root of the energy. This behavior can be understood by analyzing the electronic band structure of Al, see Fig.~\ref{fig:dos}~(b), which features a parabolic band whose minimum is located at the center of the first Brillouin zone, i.e. at the $\Gamma$ point.

\begin{figure}[htbp]
    \centering
    \includegraphics[width=0.45\textwidth]{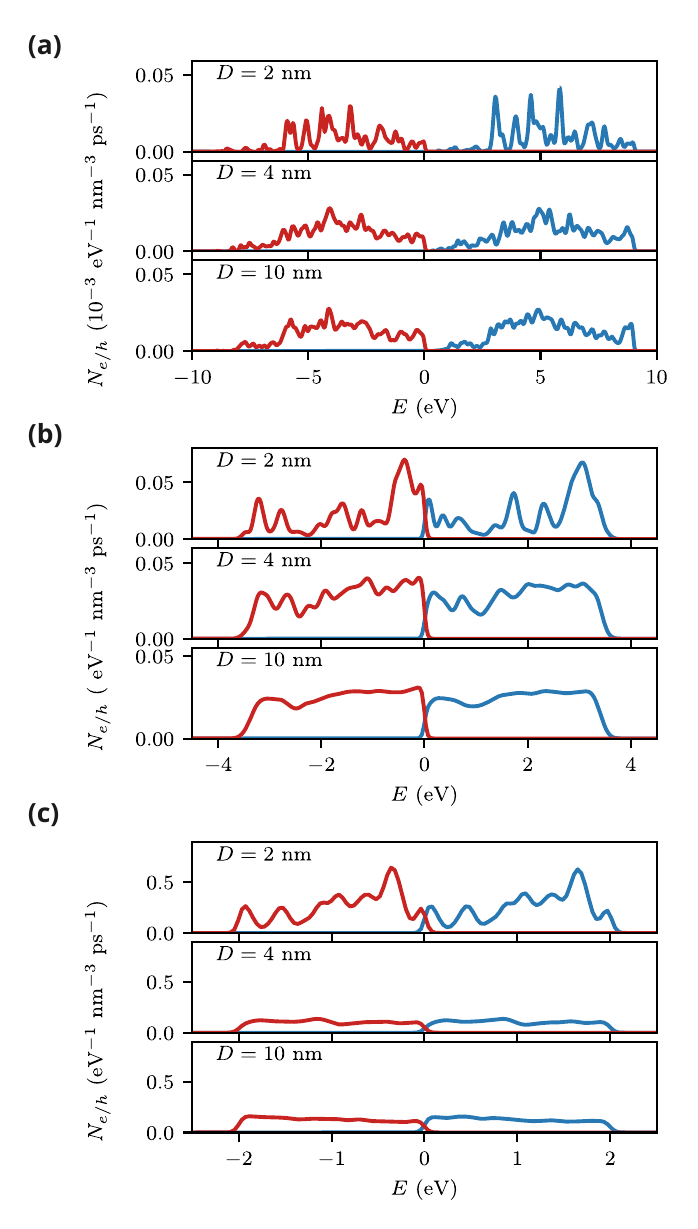} 
    \caption{Hot-carrier generation rates of spherical Al nanoparticles with different diameters $D$ in different dielectric environments. For each environment, the generation rate is calculated at the corresponding LSP energy. \textbf{(a)}: $\epsilon_m = 1$ and $\omega_{LSP}=9.0$ eV; \textbf{(b)}: $\epsilon_m = 10$ and $\omega_{LSP} = 3.4$ eV; \textbf{(c)}: $\epsilon_m = 30$ and $\omega_{LSP} = 2.0$ eV.}
    \label{fig:hot_carrier_combined}
\end{figure}

\subsection*{Hot-carrier generation}

Figure~\ref{fig:hot_carrier_combined} shows the hot-carrier generation rates as function of carrier energy of spherical AlNPs in different dielectric environments at their respective LSP energies. Similar to the DOS discussed in the previous section, we observe that the hot-carrier generation rates for small nanoparticles exhibit many discrete peaks which merge to form continuous curves for larger nanoparticles. When the nanoparticles are in vacuum (Fig.~\ref{fig:hot_carrier_combined}(a)), the LSP energy is 9.0~eV. This energy is divided between the hot electron and the hot hole. Interestingly, we find that very few electrons with energies close to the Fermi level are generated. At approximately 2 eV, a sharp increase occurs in the hot-electron generation rate and the generation rate exhibits a broad peak centered near 5 eV. As a consequence of energy conservation, the hot-hole distribution has a broad peak near -4 eV and a sharp reduction near -7 eV. 

When the dielectric function of the environment increases, the LSP energy is reduced and therefore the hot carriers are distributed over a smaller energy window around the Fermi level compared to the vacuum case, see Figs.~\ref{fig:hot_carrier_combined}(b) and (c). Notably, both the hot-electron and the hot-hole generation rate for the larger nanoparticles are almost constant over the allowed energy range. This is very different from the hot-carrier distributions of transition metal nanoparticles, such Au or Ag, which exhibit prominent peaks due to interband transitions from occupied d-states into unoccupied states with mixed sp-character.\cite{jin2020plasmon}

Our findings are in good agreement with ab initio calculations of hot-carrier generation at Al surfaces by Sundararaman and coworkers.\cite{Sundararaman2014} These authors analyzed the decay of surface plasmons due to momentum-conserving interband transitions. For small plasmon energies (2 eV), they find constant hot-electron and hot-hole distributions, while for larger plasmon energies (6 eV), no electrons with energies close to the Fermi level are generated. In addition to interband transtions, our calculations also capture LSP decay channels involving surface-enabled intraband transitions. Previous free-electron model calculations (which do not capture interband transitions) predicted that intraband transitions give rise to relatively constant hot-electron and hot-hole distributions for all nanoparticle sizes.~\cite{GOVOROV201485} This suggests that interband processes dominate LSP decay in Al nanoparticles.

\begin{figure}[htbp]
    \centering
    \includegraphics[width=1\textwidth]{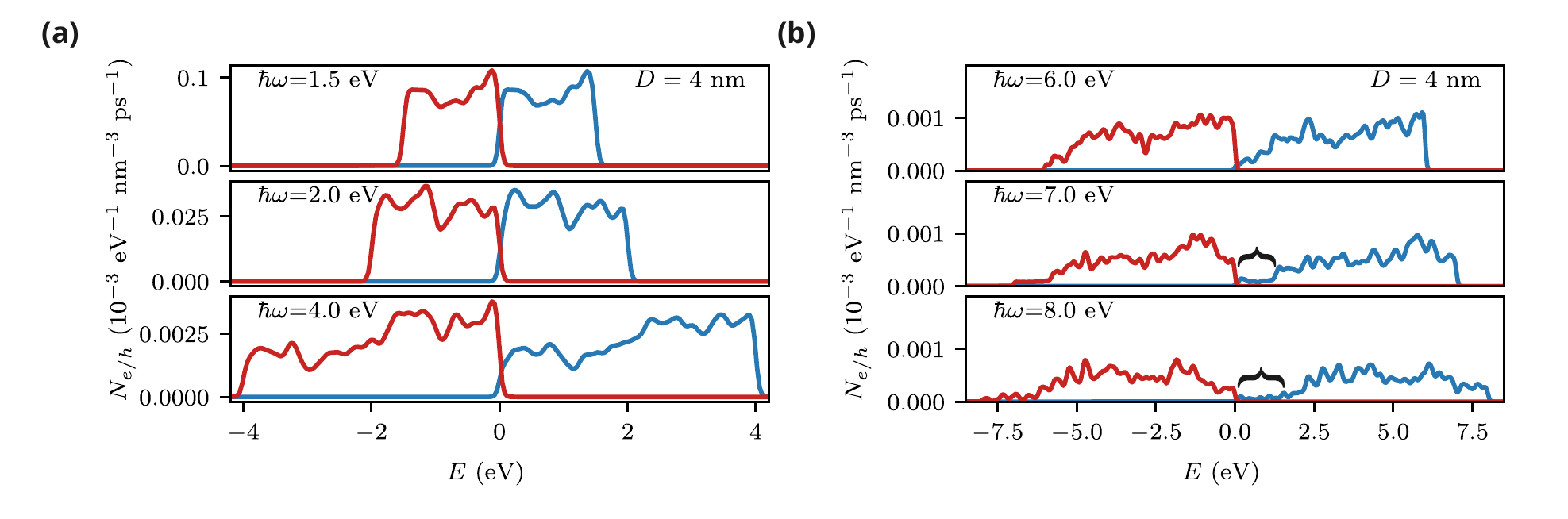} 
    \caption{Hot-carrier generation rates of spherical Al nanoparticles for photon energies of \( 1.5, 2.0, 4.0  \, \text{eV} \) \textbf{(a)} and \( 6.0, 7.0 \) and \( 8.0 \, \text{eV} \) \textbf{(b)}. Results were obtained for nanoparticles with a diameter of 4 nm, immersed in a medium with dielectric constant \( \epsilon_m \) = 30. The brackets highlight energy windows near the Fermi level in which very few electrons are generated.}
    \label{fig:hot_carrier}
\end{figure}

Finally, we study the dependence of the hot-carrier generation rate on the photon energy. Fig.~\ref{fig:hot_carrier} shows the hot-carrier generation rates of an AlNP (D = 4~nm, $\epsilon_m = 30$) for different photon energies. The lowest photon energy ($\hbar \omega = 1.5$~eV) corresponds to the low-energy shoulder in the absorption cross-section caused by interband transitions. Notably, we observe that the hot-carrier generation rates at this photon energy are significantly larger than those at the LSP energy ($\hbar\omega=2.0$~eV). This agrees with the experimental finding of Zhou et al. who observed higher rates of hot electron-induced hydrogen dissociation on AlNPs when interband transitions are optically excited compared to the excitation of LSPs.\cite{zhou:16}

While the hot-carrier generation rates are almost constant over the allowed energy range for small photon energies, they become less flat at a photon energy of 4 eV. In particular, fewer electrons with energies close to the Fermi level are generated. This trend continues at even higher photon energies, see Fig.~\ref{fig:hot_carrier}(b): at $\hbar\omega = 7$~eV, almost no electrons with energies close to the Fermi level are produced. This is consistent with the hot-carrier generation rate of nanoparticles in vacuum, see Fig.~\ref{fig:hot_carrier_combined}(a). Besides the change of their shape as a function of energy, the total magnitude of the hot-carrier generation rates becomes smaller at higher photon energies as a consequence of the weaker field enhancement when the photon energy is not resonant with the LSP energy.

\section{Conclusion}
We have studied hot‑carrier generation in spherical aluminum nanoparticles using an atomistic modeling approach. We investigated the role of nanoparticle size, incident light frequency and environment dielectric constants on the hot-carrier properties. For a range of photon energies in the visible and near-ultraviolet regime, the hot-carrier generation rates are approximately constant as function of hot-carrier energy in the allowed energy range. For higher photon energies, a reduction in the generation rate of electrons near the Fermi level is observed and attributed the band structure effects reducing the number of available interband transitions. Our calculations demonstrate the hot-carrier properties in aluminum nanoparticles are qualitatively different from those of standard plasmonic materials, such as silver or gold, and also highly tunable 
paving the way for aluminum-based nanoscale devices for energy harvesting and photonic technologies.



\begin{acknowledgement}
S.J. and J.L. acknowledge funding from the Royal Society through a Royal Society Uni-
versity Research Fellowship URF/R/191004 and also from the EPSRC programme grant
EP/W017075/1.

\end{acknowledgement}



\bibliography{achemso-demo}

\end{document}